\documentclass[%
reprint,
amsmath,amssymb,
aip,
apl,
]{revtex4-1}

\usepackage{graphicx}
\usepackage{dcolumn}
\usepackage{bm}
\usepackage{hyperref}
\usepackage{amsmath}
\usepackage{epstopdf}
\usepackage{color}
\usepackage{verbatim}
\usepackage[normalem]{ulem}
\begin{document}

\title{HgCdTe-based quantum cascade lasers operating in the GaAs phonon Reststrahlen band predicted by the balance equations method}

\author{D.V.~Ushakov}
\email{ushakovdv@bsu.by}
\affiliation{\footnotesize{\mbox{Belarusian State University, 220030 Minsk, Belarus}}}

\author{A.A.~Afonenko}
\affiliation{\footnotesize{\mbox{Belarusian State University, 220030 Minsk, Belarus}}}

\author{R.A.~Khabibullin}
\email{khabibullin@isvch.ru}
\affiliation{\footnotesize{\mbox{V.G. Mokerov Institute of Ultra High Frequency Semiconductor Electronics, Russian Academy of Sciences, 117105 Moscow, Russia}}}

\author{D.S.~Ponomarev}
\affiliation{\footnotesize{\mbox{V.G. Mokerov Institute of Ultra High Frequency Semiconductor Electronics, Russian Academy of Sciences, 117105 Moscow, Russia}}}

\author{V.Ya.~Aleshkin}
\affiliation{\footnotesize{\mbox{Institute for Physics of Microstructures, Russian Academy of Sciences, 603950 Nizhny Novgorod, Russia}}}

\author{S.V.~Morozov}
\affiliation{\footnotesize{\mbox{Institute for Physics of Microstructures, Russian Academy of Sciences, 603950 Nizhny Novgorod, Russia}}}

\author{A.A.~Dubinov}
\affiliation{\footnotesize{\mbox{Institute for Physics of Microstructures, Russian Academy of Sciences, 603950 Nizhny Novgorod, Russia}}}
\affiliation{\footnotesize{\mbox{Lobachevsky State University of Nizhny Novgorod, 603950 Nizhny Novgorod, Russia}}}

\date{\today}

\begin{abstract}
HgCdTe-based quantum cascade lasers operating in the GaAs phonon Reststrahlen band with a target wavelength of
36~$\mu$m are theoretically investigated using the balance equations method. The optimized active region designs, which are based on three and two quantum wells, exhibit a peak gain exceeding 100~cm$^{-1}$ at 150~K. We analyze the temperature dependences of the peak gain and predict the maximum operation temperatures of 170~K and 225~K for three- and two-well designs, respectively. At high temperatures ($T>120$~K), the better temperature performance of the two-well design is associated with a larger spatial overlap of the weakly localized lasing wavefunctions as well as a higher population inversion.

\end{abstract}

\maketitle

Over the past decade, the significant development of radiation sources covers a wide range of electromagnetic spectrum between mid-infrared (mid-IR) and terahertz (THz) frequencies. One of the reasons for this is a dramatic improvement of quantum cascade lasers (QCLs) performance across the far-IR and THz 
ranges \cite{2015/Vitiello/OptExpr/Quantum}. 
On the high-frequency side (10-30 THz), QCLs based on InAs/AlSb \cite{2019/Loghmari/ElectrLet/InAs-based} and InGaAs/GaAsSb \cite{2016/Ohtani/ACSPhotonics/Far-Infrared} materials have demonstrated the longest wavelengths of 25~$\mu$m and 28~$\mu$m, which correspond to frequencies of 12.0~THz and 10.7~THz, respectively. 
On the low-frequency side (1-6~THz), the highest operation frequency of 5.4 THz has been achieved by QCL based on GaAs/AlGaAs \cite{2015/Wienold/APL/Frequency}.
However, the operation of QCLs based on the above-mentioned materials in the frequency range between 6 to 10 THz is strongly affected by the optical phonon absorption in the Reststrahlen band \cite{2015/Feng/OptExpr/Photonic}. 
In particular, the extending of GaAs/AlGaAs QCL operation frequencies above 6 THz results in a significant increase of waveguide propagation loss due to a strong electron-phonon scattering in the GaAs phonon Reststrahlen band \cite{2018/Han/OptExpr/Silver, 2018/Ushakov/QE/ModeLoss}, covering 7-10 THz \cite{2016/Dyksik/JIMTW/Optical}. 
For the same reason, it is a challenge for far-IR QCLs to operate at frequencies below 12 THz because of AlAs phonon Reststrahlen band \cite{2016/Ohtani/ACSPhotonics/Far-Infrared}. Thus, one should employ alternative materials with lower or higher optical phonon energies than in GaAs, or utilize non-polar materials \cite{2019/Grange/APL/Room}, in which the interaction with optical phonons is suppressed, as materials for QCLs lasing in the GaAs phonon Reststrahlen band.

\begin{figure}[!t]
    \centering
    \includegraphics[width=0.95\columnwidth]{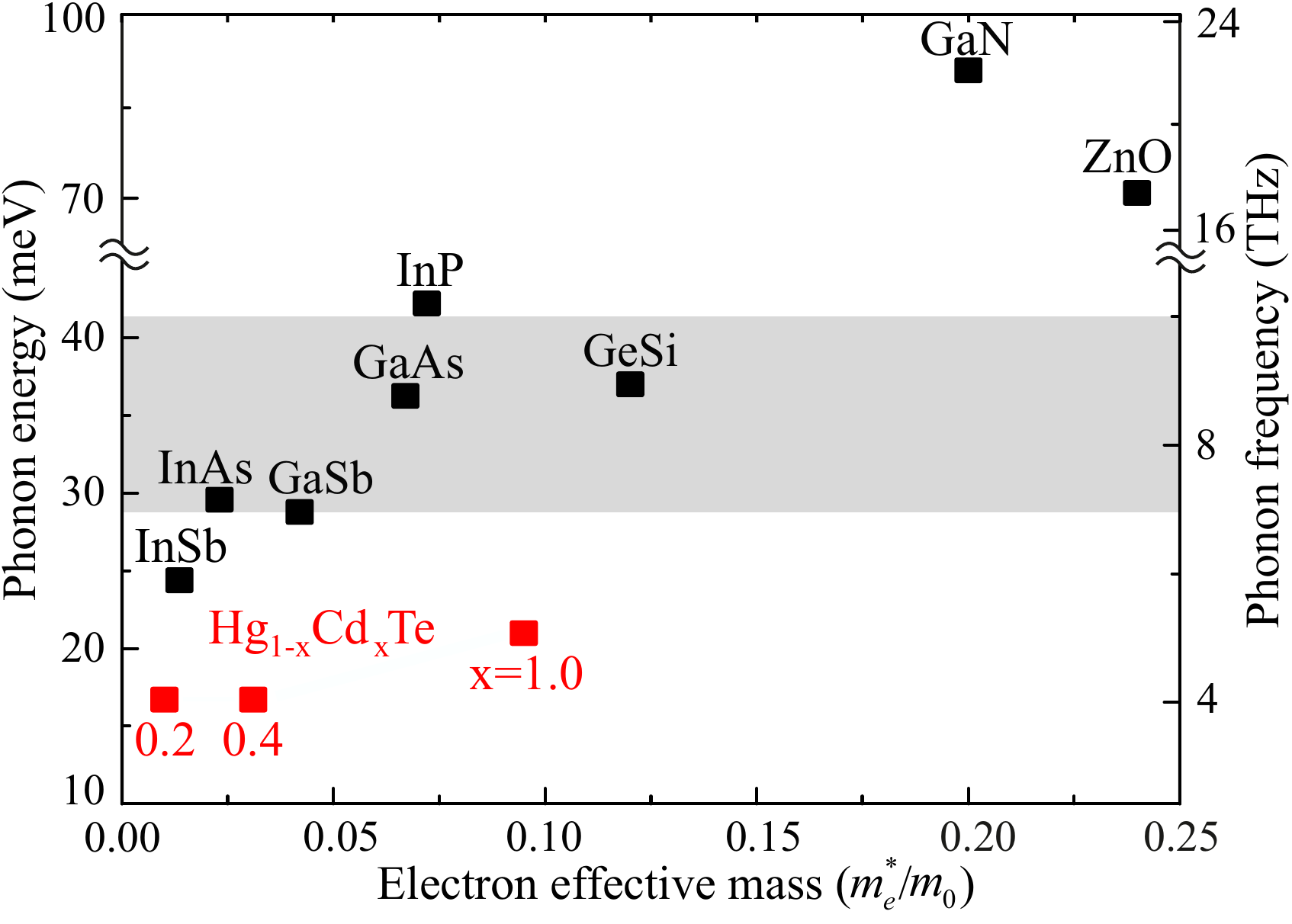}

    \quad\vspace{-0.8cm}    
        \caption{\small{
Summary of candidate materials for QCLs with corresponding electron effective mass and longitudinal optical phonon energy/frequency. The GaAs phonon Reststrahlen band is indicated by the shaded area.
}}
    \label{FIG1-m/m0}
    \quad\vspace{-0.8cm}    
\end{figure}

Semiconductors with large optical phonon energies, such as GaN or ZnO, are extremely promising materials for THz QCLs with the potential to achieve room temperature operation 
\cite{2009/Bellotti/JAP/MonteCarlo, 2012/Yasuda/JAP/Non-equilibrium}.
However, wide-bandgap materials have a higher electron effective mass $m_e^{*}$ and, therefore, a lower optical gain \cite{2008/Benveniste/APL/Influence}, since $g\sim \left(m_e^{*}\right)^{-3/2}$. Furthermore, it should be noted that low $m_e^{*}$  materials like InAs, InSb or GaSb theoretically provide higher optical amplification, but these materials are not suitable for QCLs lasing in a frequency range of 7-9~THz due to their optical phonon energy close to the GaAs one [see Fig.~\ref{FIG1-m/m0}]. In contrast to III-V semiconductors, II-VI materials, in particular, Hg$_{1-x}$Cd$_x$Te have a lower optical phonon energy than GaAs and simultaneously  low values of $m_e^{*}$. For narrow-gap Hg$_{1-x}$Cd$_x$Te with cadmium content less than 60~$\%$, the electron effective mass is comparable to conventional low $m_e^{*}$ semiconductors. Since the compound materials HgTe and CdTe are closely lattice matched (the difference in lattice constants less than 0.3~$\%$), it is possible to design a wide variety of multilayer structures based on Hg$_{1-x}$Cd$_x$Te. Moreover, there is a variety of experimentally demonstrated Hg$_{1-x}$Cd$_x$Te-based devices such as photodetectors of the mid-IR radiation (see, for example, \cite{2005/Rogalski/Reports/HgCdTe} and references therein) and coherent radiation sources operating with a wavelength of up to 19.5~$\mu$m \cite{2017/Morozov/APL/Stimulated, 2018/Fadeev/OPtExpr/Stimulated}, that shows the technological maturity of this material. Therefore, these benefits of Hg$_{1-x}$Cd$_x$Te suggest that it can be employed for QCLs operating in the GaAs Reststrahlen band.     

In this letter, to evaluate the potential of Hg$_{1-x}$Cd$_x$Te structures as a gain medium for QCL operating in the GaAs phonon Reststrahlen band with a target wavelength of 36~$\mu$m, we use the balance equation method for simulating the laser characteristics. 
In comparison with widely used approaches for modeling electronic transport in QCLs, such as the nonequilibrium Green's function formalism \cite{2003/Lee/APL/Theoretical} or the density matrix formalism \cite{2017/Jirauschek/JAP/Density}, the balance equation method allows a significant reduction of the computation time. 
Thus, we have an opportunity to consider a wide variety of active region designs, as well as optimize the given design by varying of the thickness of all layers in the wide range. 
Moreover, the balance equation method allows us to account the dephasing effect, which clearly plays an important role in QCLs  \cite{2005/Callebaut/JAP/Importance,2009/Kumar/PhysRevB/Coherence}, and predicts the formation of electric field domains across the active region \cite{2019/KHABIBULLIN/OptoElRev/TheOperation}. 
Previously, we have demonstrated the good agreement between calculated and experimental light-current-voltage characteristics for the GaAs/AlGaAs QCL emitting near 2.3~THz~\cite{2019/Ushakov/QE/Balance-equation}. 
In case of HgCdTe QCLs, it is important to take into account the strong effect of conduction band non-parabolicity in the calculation of electronic states, which is done using the 3-band k$\cdot$p Hamiltonian \cite{1994/Sirtori/PhysRevB/Nonparabolicity}. 
In addition, an anomalous temperature dependence of the Hg$_{1-x}$Cd$_x$Te band gap \cite{2005/Rogalski/Reports/HgCdTe} is included in our model.

Our choice of the Hg$_{1-x}$Cd$_x$Te/Hg$_{1-y}$Cd$_{y}$Te composition for quantum wells/barriers is motivated by the following reasons. 
On the one hand, the increase of mercury content leads to a decrease in the $m_e^{*}$, which is favorable from a theoretical perspective. 
On the other hand, Hg$_{1-x}$Cd$_x$Te with a Hg composition more than 80~$\%$ becomes gapless material with a high interband absorption at the target wavelength as well as with a detrimental effect of mixing between the states of electrons and holes \cite{1999/Paula/PRB/Interband}. 
Thus, we have used the Hg$_{0.8}$Cd$_{0.2}$Te wells as an optimum trade-off between the above-mentioned tendencies. 
The height of Hg$_{1-y}$Cd$_y$Te barriers is defined as a compromise between the suppression of parasitic leakage current and the difficulties of growth technology. 
In order to prevent the current leakage into the continuum it is necessary to reduce the Hg content for a higher potential barrier. 
However, an increase in the conduction band offset results in a decrease in the barrier thickness, creating high requirements on the epitaxial growth. 
Thus, we prefer Hg$_{0.8}$Cd$_{0.2}$Te/Hg$_{0.6}$Cd$_{0.4}$Te material system, which provides $\sim$~289~meV conduction band offset at 50~K and allows designing an active region of QCLs with technological maturity.

We analyze three- and two-well designs with resonant-phonon depopulation scheme when the operation bias per period is close to the sum of the emission photon energy $\hbar w =34.3$~meV and the longitudinal optical phonon energy 17.8~meV. 
Our choice of the numbers of wells in the period is motivated by the highest operation temperatures of THz QCLs based on GaAs/AlGaAs with 3-well \cite{2012/Fathololoumi/OptExpr/Terahertz} and 2-well \cite{2019/Bosco/APL/Thermoelectrically} designs. 
The balance equation method allows us to optimize the chosen HgCdTe-designs in order to achieve the maximum gain for emission frequency of 8.3~THz, varying all barrier layers thicknesses from 1 to 8~nm and wells thicknesses from 3 to 30 nm with a calculation step equal to half the lattice constant of CdTe $\sim 3.25$~\AA.

\begin{figure}[!t]

\noindent
    \includegraphics[width=0.95\columnwidth]{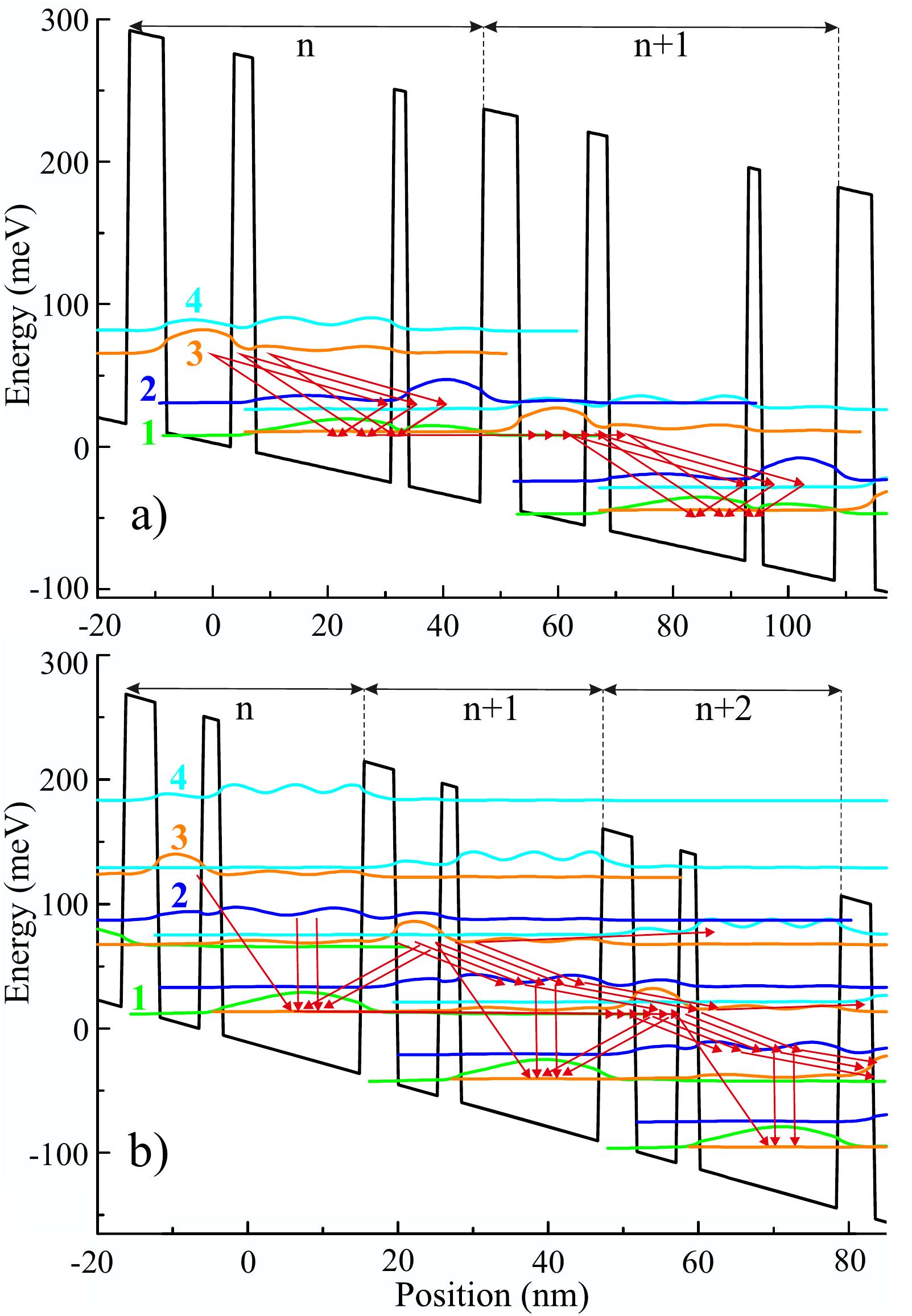}
   
    \caption{\small{
        Conduction band diagram and squared modules of wavefunctions for the Hg$_{0.6}$Cd$_{0.4}$Te/Hg$_{0.8}$Cd$_{0.2}$Te QCLs computed by 3-band Hamiltonian k$\cdot$p-method at a bias of 55~mV/period are demonstrated for (a) two neighboring periods of 3-well design at 100~K and 
    (b) three neighboring periods of 2-well design at 200~K. The red arrows indicate the main current channels through the structure, taking into account that the total current through the period equals to 6 red arrows for 3-well design and 9 red arrows for 2-well design.}}
    
    \quad\vspace{-0.6cm}    
    \label{FIG-3QW-2QW-Ec}
\end{figure}

As a result of numerical optimization, we have found the 3-well design with peak gain exceeding of 100~cm$^{-1}$ at a lattice temperature of 150~K. 
The layer sequence of the single period of optimized design in nanometers 
{\bf 6.5}/11.7/{\bf 3.9}/\underline{24.0}/{\bf 2.6}/13.0 with Hg$_{0.6}$Cd$_{0.4}$Te barriers indicated in bold letters and Hg$_{0.8}$Cd$_{0.2}$Te wells, whereas the central part of underlined well is doped with sheet electron density of $6.2\cdot 10^{10}$cm$^{-2}$. 
The band structure and the current flow through the energy levels as simulated by the balance equation method are shown in Fig.~\ref{FIG-3QW-2QW-Ec}~(a). 
The operation principle of the present design is similar to the conventional 3-well THz QCLs, when the population inversion is based on electron injection into the upper laser level 3 of $n+1$ period via resonant tunneling from injector level 1 of $n$ period  and electron depopulation of the lower laser level 2 of $n+1$ period occurs with resonant phonon emission landing on the injector level 1 of the same period.
This principle is well illustrated by the calculated main current channels through the structure, including the parasitic current channel between the upper laser level 3 and the injector level 1, which are indicated by red arrows. 
 
The best temperature robustness is demonstrated by the optimized 2-well design with peak gain exceeding of 100 cm$^{-1}$ at 200~K and layer sequence (in nanometers with barriers in bold letters) {\bf 4.5}/5.8/{\bf 2.6}/\underline{18.8} with a doping density of $3.2\cdot 10^{10}$ cm$^{-2}$. 
As compared with the 3-well design, narrower wells in the 2-well design lead to higher energy levels with weakly localized electron wavefunctions spanning over several periods [see Fig.~\ref{FIG-3QW-2QW-Ec}~(b)]. Consequently, the 2-well design has a larger spatial overlap of wavefunctions of the laser levels 3 and 2, which increases the dipole matrix element of the radiation transition. On the other hand, utilizing high energy levels in the 2-well design has two detrimental effects. First, the effective mass of the above levels becomes higher due to the non-parabolic band effect. Second, the electron leakage is activated on high energy levels by the parasitic current channel $3 \rightarrow 4$ with sequential tunneling into the continuum.

In the 2-well design the trajectory of the current flow becomes more complex with the main ''stream'' through laser levels 3 and 2 and the minor current channel based on diagonal transitions between levels 3 and 1, which occur in the direction and opposite to the direction of the current flow. Furthermore, the present design is characterized by the hybrid injection scheme based on resonant-tunnelling transitions from injector level 1 of $n$ period to upper laser level 3 of $n+2$ period and scattering-assisted injection from the lower laser level 2 of $n$ period to the upper laser level 3 of $n+1$ period with resonant emission of optical phonon. We emphasize that in the 2-well design the population of the lower laser level 2 is less than the population of the upper laser level 3, since the populations of these levels are related to each other by the Bolzmann distribution $n_2/n_3 \approx \exp \left(  - \hbar \omega_{\rm{LO}}/kT \right)$. 
Finally, in the 2-well design the parasitic injection from level 1 to the lower laser level 2 is suppressed due to the minimized overlap between wavefunctions of the corresponding levels. For these reasons, we expect that the 2-well design is less sensitive to the temperature and allows achieving higher operation temperatures than the 3-well design.

\begin{figure}[!t]

        \includegraphics[width=0.95\columnwidth]{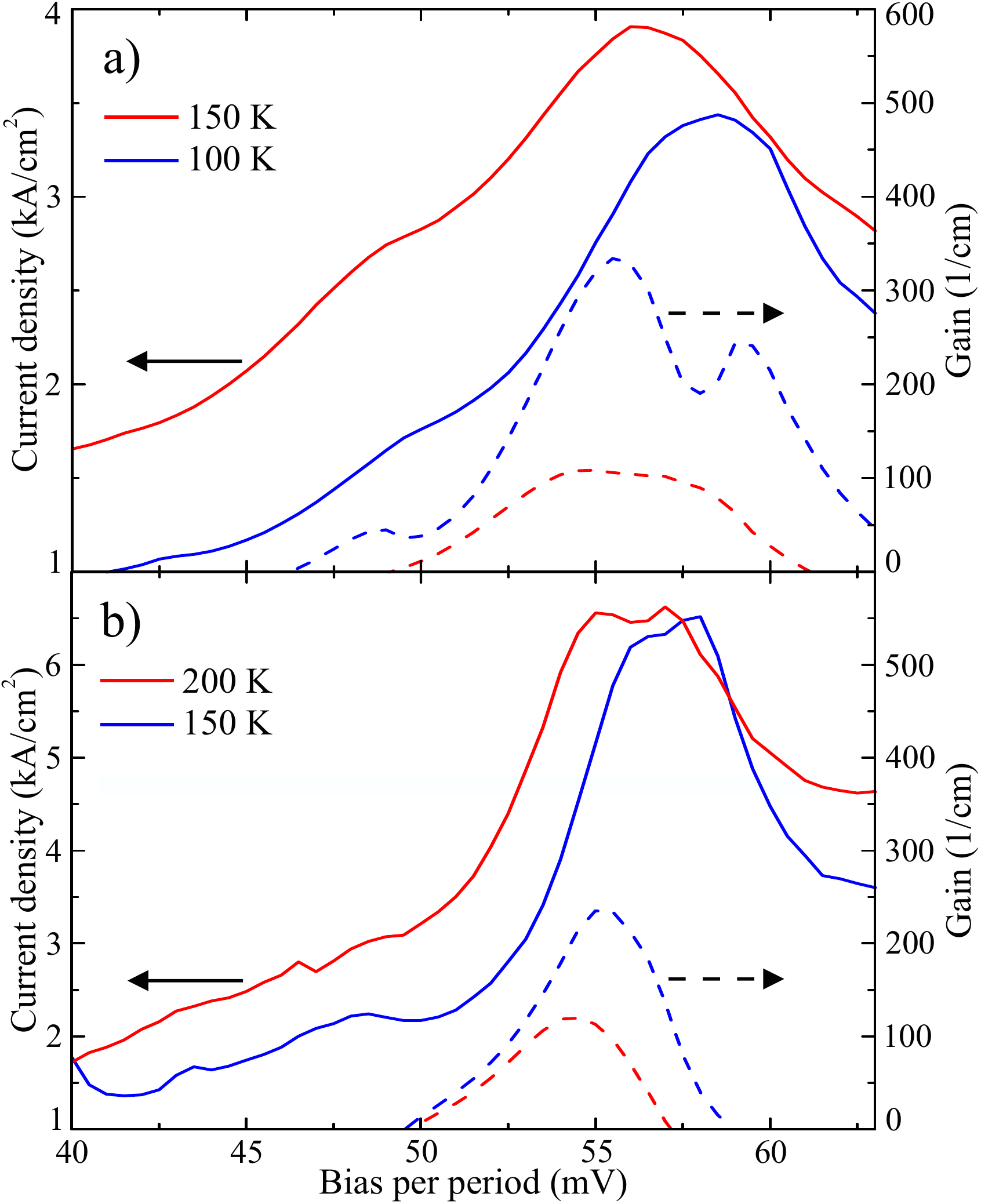}

       \quad\vspace{-0.8cm}    
 \caption{\small{
The current density (solid lines) and peak gain (dashed lines) versus voltage are calculated using the balance equations method for (a) 3-well design at a lattice temperatures of 100 K and 150 K and (b) 2-well design at a lattice temperatures of 150 K and 200 K. 
}}
    \label{FIG-3QW-2QW-j-g-V1}
       \quad\vspace{-0.8cm}
       \end{figure}

In Fig.~\ref{FIG-3QW-2QW-j-g-V1}~(a), we compare the calculated current density-voltage and peak gain-voltage characteristics of the optimized 3-well design at two temperatures of 100~K and 150~K. 
One observes that the operation bias point near 55~mV/period, where the peak gain has a maximum value, is below the peak-current bias at both temperatures. 
Thus, the present design demonstrates a smooth current-voltage characteristic with operation bias point outside the region of negative differential resistance to prevent electrical instability. 
It is worth noting that this design has a strong temperature dependence of the current density, which leads to its increasing from 2.8 to 3.8~kA/cm$^2$ at operation bias point as temperature increases by 50~K. This temperature effect is associated with a $\sim20~\%$ increase of the Bose factor of the electron-phonon scattering rate due to a low optical phonon energy in HgCdTe. In addition, the population of the injector level 1 decreases by 10~\%, because the populations of the lower laser level 2 and the high energy level 4 increase by 6~\% and 4~\%, respectively. As a result, new parasitic channels $4 \rightarrow 1$ and $4 \rightarrow 2$ of current leakage occur at high temperatures. Moreover, with increasing temperature from 100~K to 150~K, the peak gain rapidly drops and the operation bias range shrinks. Thus, the optimized 3-well design depends significantly on temperature and is more suitable for low temperature operation.

The results of the calculated current density and peak gain versus applied voltage at 150 K and 200 K for the 2-well design are shown in Fig.~\ref{FIG-3QW-2QW-j-g-V1}~(b). As well as for the 3-well design, the operation bias point near 55 mV/period is on the increasing branch of the current-voltage characteristic. However, we observe that the local gradient of the current-voltage curve of the 2-well design is larger than in the 3-well design. Reasonably, high energy levels of the 2-well design are more sensitive to the applied bias, resulting in a high differential resistance and a narrower operation bias range. It should also be noted that the current density in the 2-well design is two times higher than in the 3-well design. We attribute this to the presence of electron leakage in the 2-well design from high energy levels into the continuum.

\begin{figure}[!b]

     \includegraphics[width=0.95\columnwidth]{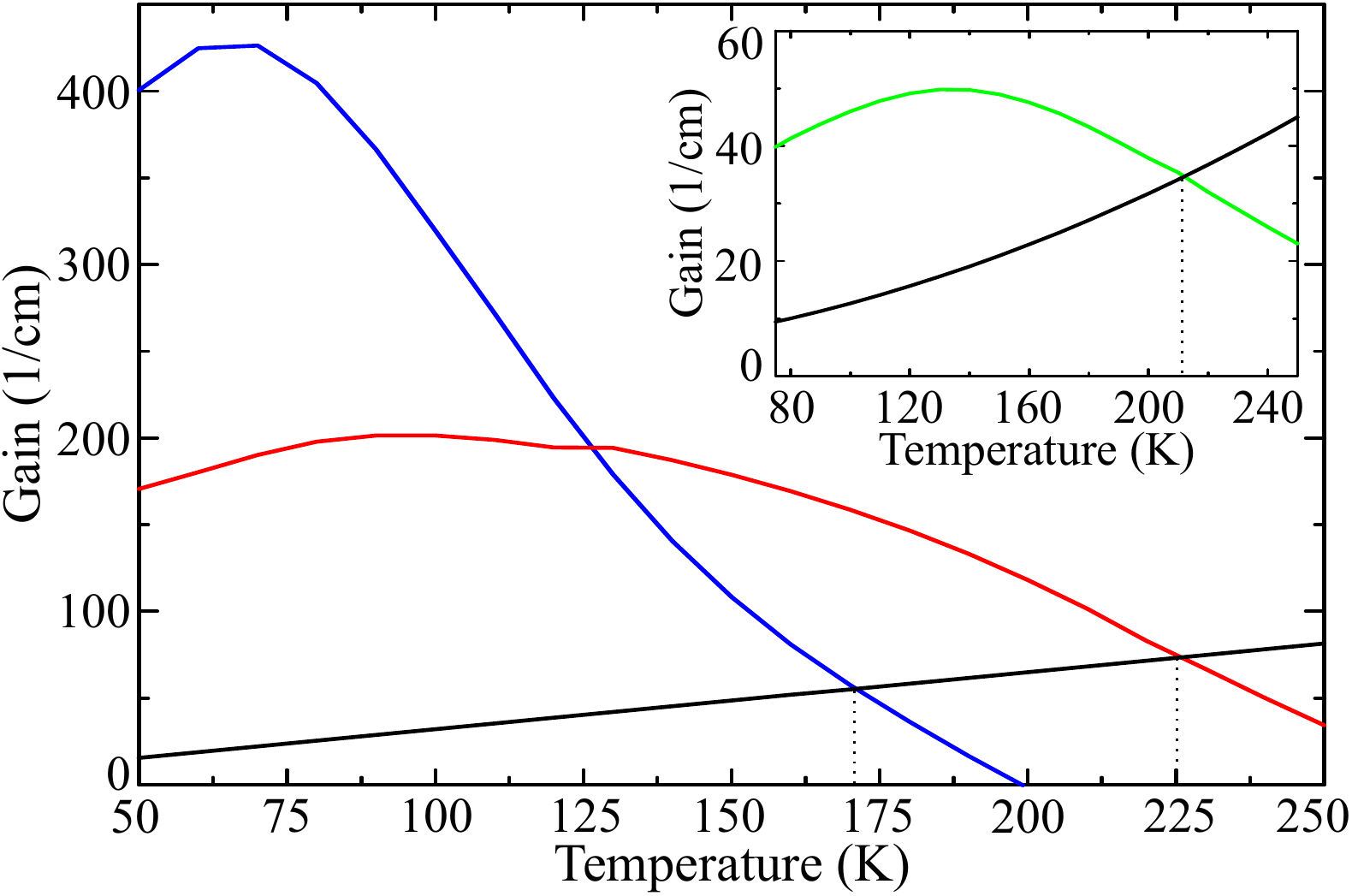}

    \caption{\small{
        The temperature dependencies of peak gain are shown for the HgCdTe designs based on 3-well (blue line) and 2-well (red line), and for the GaAs/AlGaAs design based on 2-well of Bosco et al. \cite{2019/Bosco/APL/Thermoelectrically} (green line in the inset). Cavity losses of a Cu-Cu waveguide is indicated by the black curves.  
     }}
    \label{FIG-3-2QW-G}
           \quad\vspace{-0.8cm}
\end{figure}

In Fig.~\ref{FIG-3-2QW-G}, we compare the temperature dependence of the peak gain at 8.3~THz of the optimized 2-well and 3-well designs. 
At low temperatures, the 3-well design demonstrates the highest peak gain, which begins to decrease rapidly at temperatures above 70~K. 
Conversely, in case of the 2-well design, the peak gain has lower values with weak temperature dependence up to 150~K and gradually decreases at higher temperatures. 
This leads to significantly higher peak gain of the 2-well design at temperatures above 120~K and, as a result to higher operating temperatures. 
In order to estimate the maximum operation temperatures of the given designs, we calculate the cavity losses for a 12~$\mu$m thick Cu-Cu waveguide based on HgCdTe as a function of temperature using the method proposed in 
Ref.~\cite{2018/Ushakov/QE/ModeLoss}. In this case, we take into account scattering by optical phonons \cite{1972/BAARS/SSC/Reststrahlen, 1974/Grynberg/PhysRevB/Dielectric} and free charge carriers \cite{1983/Mroczkowski/JAP/Optical} for HgCdTe QCL with 75~nm and 50~nm thick $n^+$-CdTe contact layers with a doping concentration of $10^{17}$~cm$^{-3}$. Now we may clearly identify the temperature, which ceases the laser action. Under this temperature the computed optical gain becomes equal to the cavity losses. Such conditions are achieved for the 2-well and 3-well designs at 225~K and 170~K, respectively, which corresponds to the maximum operating temperature of the given designs. To validate the predictability of our approach, we calculate the above dependencies for the 2-well design of GaAs/AlGaAs QCL emitting near 3.9~THz with the highest operation temperature of 210.5~K achieved to date \cite{2019/Bosco/APL/Thermoelectrically}. Finally, we find excellent agreement between the predicted maximum operating temperature of 212 K [see inset of Fig.~\ref{FIG-3-2QW-G}] and the experimental value.

\begin{figure}[!t]

        \includegraphics[width=0.95\columnwidth]{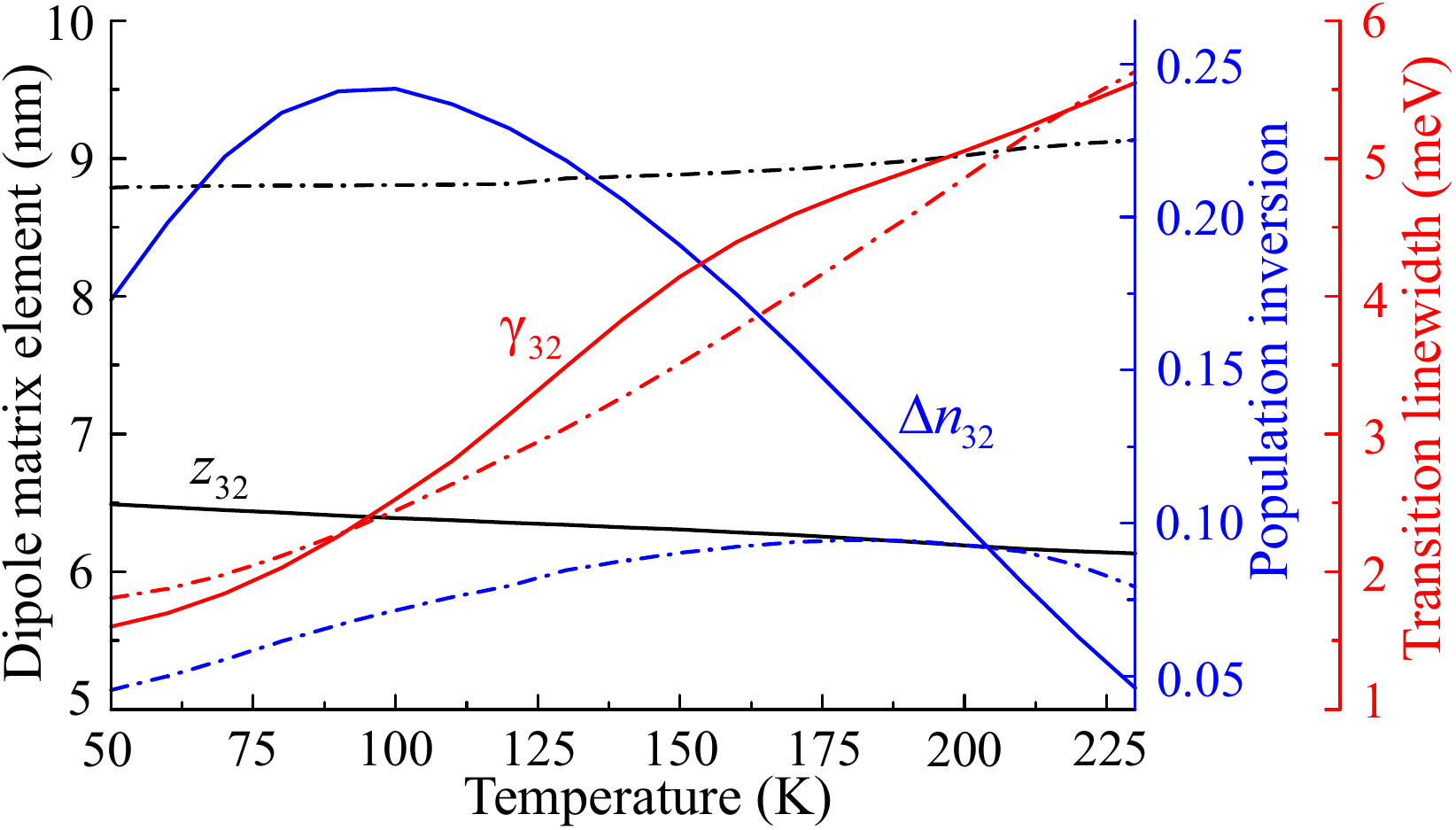}
        
    \caption{\small{
      The temperature dependencies of dipole matrix element $z_{32}$,  population inversion $\Delta n_{32}$  and transition linewidth $\gamma_{32}$ are shown for the optimized 3-well (solid curves) and 2-well (dash-dotted curves) designs.
     }}
    \label{FIG-3-2QW-N-Z-gamma}
           \quad\vspace{-0.8cm}
\end{figure}

To highlight the difference in the temperature performance of the 2-well and 3-well designs, 
in Fig.~\ref{FIG-3-2QW-N-Z-gamma} we analyze the main parameters defining the peak gain for both designs as $g_{32} \sim \Delta n_{32}|z_{32}|^2/\gamma_{32}$, where $\Delta n_{32}$  is the population inversion,  $z_{32}$ is the dipole matrix element, and $\gamma_{32}$ is the transition linewidth. 
Despite of the low value of $\Delta n_{32}$ for the 2-well design, this parameter slightly increases at elevated temperatures and becomes higher than the $\Delta n_{32}$ of the 3-well at temperatures above 200 K. One more reason of high temperature operation of the 2-well design is a higher value of  $z_{32}$, which increases from 8.8~nm to 9.2~nm with increasing temperature from 50 to 250~K. Meanwhile the 3-well design demonstrates a lower value of  $z_{32}$, which is reduced from 6.3~nm to 6.1~nm within the above temperature range. Thus, the 2-well design has a higher $\Delta n_{32}\cdot z_{32}$ product than the 3-well design at temperatures above 200 K with close $\gamma_{32}$ for both designs, leading to a higher peak gain and, consequently, to a higher operation temperature.

In conclusion, we have demonstrated the potential of the HgCdTe structures as a gain medium for QCL operating in the GaAs phonon Reststrahlen band. Carrier transport and optical gain properties are theoretically investigated using the balance equations method with a 3-band $k\cdot p$ Hamiltonian. The HgCdTe active region is optimized by varying of the layer thicknesses in the wide range in order to achieve the maximum gain for emission frequency of 8.3 THz. Finally, we have proposed the three- and two-well designs with maximum operation temperatures of 170 K and 225 K, respectively. We believe that the HgCdTe structures may be a promising candidate for QCLs operating in the frequency range from 7 THz to 10 THz.

This work was supported by the Belarusian republican foundation for fundamental research Grant No.~F18R-107 and the Russian foundation for basic research Grant No.~18-52-00011\underline{ }Bel in the part of band diagram calculations, and the Russian Science Foundation Grant No.~18-19-00493 in the part of HgCdTe QCLs modelling.

\bibliography{THz_HgCdTe_REF}

\end{document}